\documentclass[dvipsnames,aps,pre,nobibnotes,twocolumn,superscriptaddress,nolongbibliography]{revtex4-2}
\usepackage[T1]{fontenc}

\usepackage[unicode=true,pdfusetitle,
 bookmarks=true,bookmarksnumbered=false,bookmarksopen=false,
 breaklinks=false,pdfborder={0 0 0},pdfborderstyle={},backref=false,colorlinks=true]
 {hyperref}
 \hypersetup{
 citecolor=blue,
 urlcolor=blue,
 linkcolor=blue}
\usepackage{soul}
\usepackage[utf8]{inputenc}
\usepackage{latexsym,amsmath,verbatim}
\usepackage{amssymb} 
\usepackage{amsfonts}
\usepackage{colortbl}
\usepackage{array}
\usepackage{stackrel}
\usepackage{bm}
\usepackage{nicefrac}
\usepackage{rotating}
\usepackage[final]{pdfpages}
\usepackage{lipsum}
\usepackage{centernot}

\makeatletter
\AtBeginDocument{\let\LS@rot\@undefined}
\makeatother

\begin{document}

\title{Non-normal Dynamics on Non-reciprocal Networks: \\
Reactivity and Effective Dimensionality in Neural  Circuits}


\author{Anna Poggialini}
\email{anna.poggialini@uniroma1.it}
\affiliation{Dipartimento di Fisica Universit\`a ``Sapienza'', P.le
  A. Moro, 2, I-00185 Rome, Italy.}
\affiliation{`Enrico Fermi' Research Center (CREF), Via Panisperna 89A, 00184 - Rome, Italy.}
\affiliation{Dipartimento di Scienze Biomediche, Universit\`a di Padova, Padova 35131, Italy.}
\author{Serena Di Santo}

\affiliation{Departamento de Electromagnetismo y F\'isica de la Materia, Universidad de Granada, Granada 18071, Spain}
\affiliation{Instituto Carlos I de F\'isica Te\'orica y Computacional, Univ. de Granada, E-18071, Granada, Spain.}
\author{Pablo Villegas}
\affiliation{`Enrico Fermi' Research Center (CREF), Via Panisperna 89A, 00184 - Rome, Italy}
\affiliation{Instituto Carlos I de F\'isica Te\'orica y Computacional, Univ. de Granada, E-18071, Granada, Spain.}
\author{Andrea Gabrielli}
\affiliation{Dipartimento di Ingegneria Civile, Informatica e delle Tecnologie Aeronautiche, Universit\`a degli Studi ``Roma Tre'', Via Vito Volterra 62, 00146 - Rome, Italy.}
\affiliation{`Enrico Fermi' Research Center (CREF), Via Panisperna 89A, 00184 - Rome, Italy}
\affiliation{Istituto dei Sistemi Complessi (ISC) - CNR, Rome, Italy.}
\author{Miguel A. Mu\~noz}
\affiliation{Departamento de Electromagnetismo y F\'isica de la Materia, Universidad de Granada, Granada 18071, Spain}
\affiliation{Instituto Carlos I de F\'isica Te\'orica y Computacional, Univ. de Granada, E-18071, Granada, Spain.}

\begin{abstract}
Non-reciprocal interactions are widespread in neural circuits, arising both from asymmetric local excitation–inhibition dynamics and from directed coupling between distinct populations. Here we investigate how these two sources of asymmetry jointly shape network activity using a modified Wilson-Cowan framework in which each node is an excitatory/inhibitory population operating in a strongly non-normal regime and nodes are linked by directed excitatory projections. We analyze minimal three-node motifs that interpolate between a feedback (cyclic) architecture and a purely feedforward chain. By combining steady-state and bifurcation analysis with linear stability and reactivity measures, we show that feedback coupling largely maps onto an effective parameter shift relative to the uncoupled case, whereas feedforward coupling produces qualitatively different activation properties, including distinct scaling across stages and a genuine enhancement of reactive transient amplification. Incorporating intrinsic finite-size noise, we find fluctuation-driven switching between inactive and active states and motif-dependent spatiotemporal organization: feedback promotes coherent, correlated activation across nodes, while feedforward networks display sequential, step-like propagation. Together, these results provide a minimal framework for understanding how multi-scale non-reciprocities control transient responses and noise-driven dynamics in coupled neural populations.
\end{abstract}
\maketitle

\section{Introduction}

Non-reciprocity has recently emerged as a powerful organizing principle in the study of complex systems far from equilibrium. It describes the asymmetric, directed nature of interactions between components, whereby the influence of unit A on unit B is not necessarily matched by the reverse influence of B on A. This asymmetry underlies many forms of rich and often unexpected behavior across a broad range of domains, from active matter and synthetic meta-materials to neural circuits and ecological webs \cite{Fruchart2021, Coulais2017, Martorell2024}. Non-reciprocal interactions break detailed balance and time-reversal symmetry at the microscopic level, enabling new types of collective dynamics ---including, among others, spontaneous oscillations, traveling waves, and fluctuation-amplified responses--- that do not exist in symmetric, equilibrium-like systems \cite{Nicoletti2018, Hennequin2012, Asllani2018, Loos2020}.

In this paper, we distinguish two manifestations of physical non-reciprocity in complex systems—arising either from asymmetric local excitatory-inhibitory dynamics at a local level or from directed network architecture—which often co-occur but are conceptually separate. We use   non-normal   linear operators as a convenient mathematical framework to describe both types of non-reciprocity and 
analyze their dynamical consequences. 
Recall that a matrix is non-normal when it does not commute with its transpose (for real matrices) or adjoint (for complex matrices), e.g. $AA^{\top}\neq A^{\top}A$. In such cases, 
its (right) eigenmodes are generally non-orthogonal, and this non-orthogonality can enable strong   transient amplification   even when all eigenvalues indicate asymptotic stability. 

The first form we address is \emph{local dynamical non-normality}, which arises at the level of the internal dynamics of individual units or subsystems. For example, consider an island with a predator population and a prey population. The local Jacobian (stability matrix) around a stationary point of the dynamics can be non-normal, so that even if the fixed point is asymptotically stable, the system may exhibit large transient responses to small perturbations—an effect that can be enhanced by continual disturbances or noise \cite{Neubert, Neubert2002, Trefethen, Livi2019,  Troude2025a, Troude2025b}. Similarly, in neural circuits, interacting excitatory and inhibitory  neuronal populations often operate in regimes where the effective linearized dynamics is strongly non-normal, allowing transient amplification of input fluctuations and supporting fast signal processing, selectivity, and flexible population coding \cite{Hennequin2014, Asllani2014, Asllani2018, Asllani2018top, Muolo2020, DiSanto2018, Buendia2019, Wang-Sornette,Baggio2021}.

The second form is what we term \emph{structural non-normality}, which stems from network-level asymmetry in how units influence one another at a larger scale. In many real-world systems, especially biological, and engineered networks, the connectivity that mediates inter-unit interactions is directed and typically not symmetric. Ecological and social systems, for example, often exhibit trophic or influence hierarchies, in which some quantity flows preferentially along directed pathways \cite{Virginia-PNAS, Johnson}. Neural circuits likewise are not fully bidirectional: higher-order cortical areas send feedback to early sensory regions, but not necessarily with equal strength in the opposite direction \cite{Gilbert2013, Friston2005, Semedo2022,Hennequin2012}. Such directed structure can (but need not) generate a non-normal connectivity operator; when it does, it can promote directional amplification, long-range correlations, wave-like responses, and flow-like propagation of activity across the network.

Here, we test the hypothesis that embedding locally non-normal units within globally non-reciprocal architectures yields collective behaviors stronger or  distinct from those produced by either source of asymmetry alone. Specifically, we explore how the interplay between local dynamical non-normality (within E-I populations) and global structural non-normality (across network tiers) determines the reactivity of the system. We show that while feedback architectures largely rescale the effective parameters of local units, feedforward motifs fundamentally alter the scaling laws of the system, introducing a hierarchy of timescales and amplifying transient responses. In other words, when locally asymmetric (and potentially non-normal) subsystems (e.g., excitatory/inhibitory microcircuits) are embedded in a directed large-scale architecture (e.g., feedforward/feedback pathways), the resulting macroscopic dynamics may exhibit emergent phenomena not captured by standard linear-response intuition for symmetric interactions. These may include enhanced stochastic amplification, novel transient modes, and selective propagation of fluctuations and signals across the network.

As a test of this hypothesis, here we analyze a minimal yet general neural model consisting of multiple interacting dynamical units, each described by a non-normal local E-I system, coupled through a network, that in general is non-reciprocal. Using tools from linear stability analysis, stochastic dynamics, and spectral theory, we explore how the composition of non-reciprocities ---dynamical and structural--- governs the system’s susceptibility to noise, its capacity to transmit signals, and its repertoire of emergent dynamical states.

While our focus is primarily on neural systems ---where both forms of non-reciprocity are particularly notorious--- the conceptual framework we propose is broadly applicable to other domains, including active materials, biological networks, and collective decision-making systems. In all these settings, non-reciprocity functions as a structural and dynamical source of complexity, and elucidating its distinct manifestations is crucial for understanding the behavior of high-dimensional, complex systems.

\section{Wilson-Cowan model for neural populations}

\subsection{Dynamical non-normal effects in E/I populations.}

We consider the classical Wilson-Cowan model \cite{Wilson1972}, which describes the dynamics of a large neural population composed of homogeneously interconnected excitatory and inhibitory neurons \cite{WC2016,WC2021,Benayoun2010,DiSanto2018}. The system of differential equations governing the dynamical evolution of the model is given by:
\begin{align}
\begin{aligned}
\frac{dx(t)}{dt} &= -\alpha x(t) + (1 - x(t)) f(s) \\
\frac{dy(t)}{dt} &= -\alpha y(t) + (1 - y(t)) f(s),
\end{aligned}
\label{Eq.s=}
\end{align}
where $x(t)$ and $y(t)$ represent the levels of  activation of excitatory and inhibitory neuronal population, respectively. The parameter $\alpha$ represents the rate of spontaneous activity decay, while $s$ denotes the net incoming input,
$s(t) = \gamma_{\mu}x(t) - \gamma_{\nu} y(t) + h $,
which is simply the sum of all synaptic excitatory and inhibitory inputs, weighted by their respective synaptic strengths   plus a  constant external input, $h$. 
To reduce the dimensionality of the parameter space, all synaptic weights of a given type have been set to the same value:
$(\gamma_{\mu}$ for excitation and $\gamma_{\nu}$ for inhibition). Finally, $ f(s)$ is a sigmoid non-negative response function.
The autonomous system ($h=0$) displays two distinct dynamical regimes: a quiescent (inactive) phase, where 
the mean activity vanishes,
$\bar{x} = \bar{y} = 0$, and an active phase characterized by nonzero steady-state activity  $\bar{x} = \bar{y} > 0$. These phases are separated by a bifurcation that occurs at the critical point $\gamma_{\mu} = \gamma_{\nu} + \alpha$: below this threshold, the system remains in the quiescent state, while above it, activity becomes self-sustained (see below). 

It is useful to make the change of variables $\Sigma = (x + y)/2$ and $\Delta = (x - y)/2$, to consider the overall activation and the differential one between excitation and inhibition, respectively. 
Linearizing around the active-state fixed point $(\bar\Sigma, \bar\Delta \neq 0)$, one obtains a strictly upper-triangular Jacobian matrix, which is manifestly non-normal:
\begin{equation}
J = 
\begin{pmatrix}
-\lambda_1 & \omega_{\text{ff}} \\
0 & -\lambda_2
\end{pmatrix},
\end{equation}
where the eigenvalues are given by $\lambda_1 = \alpha + f(\bar s) + (1 - \bar\Sigma) \omega_0 f'(\bar s)$ and $\lambda_2 = \alpha + f(\bar s)$, with $\bar s = \omega_0 \bar\Sigma + h$. The feedforward term $\omega_{\text{ff}} = (1 - \bar\Sigma)(\gamma_\mu + \gamma_\nu) f'(\bar s)$ governs the asymmetric influence of  $\Sigma$ on $\Delta$, while the parameter $\omega_0 = \gamma_\mu - \gamma_\nu$ quantifies the difference between excitatory and inhibitory couplings. A key feature of this structure is that the eigenvectors of $J$ are non-orthogonal  \cite{Benayoun2010, DiSanto2018} and tend to become nearly parallel under the 
condition $\zeta \equiv \frac{|\gamma_\mu - \gamma_\nu|}{\gamma_\mu + \gamma_\nu} \approx 0$, which describes balance between excitation and inhibition. In this regime, the eigenbasis poorly spans the full phase space: directions orthogonal to the dominant eigenvectors are poorly represented, and any perturbation with a component in these directions requires large contributions from the available (nearly aligned) eigenvectors. As a result, even small fluctuations along $\Delta$ can lead to giant transient excursions in $\Sigma$, despite the presence of negative eigenvalues and overall linear stability \cite{Benayoun2010, DiSanto2018, Corral}. This reflects the hallmark of non-normal dynamics: strong directional sensitivity and transient amplification that escape standard eigenvalue-based stability analysis.

This type of structure has important consequences, especially in the presence of stochasticity. In fact, Benayoun \emph{et al.} introduced stochastic effects into the Wilson–Cowan framework by deriving it from a microscopic model with a finite population of $N$ neurons \cite{VanKampen_Book,Gardiner1985_book,Benayoun2010}. This approach leads to a system of coupled Langevin equations ---interpreted in the It{o} sense--- that captures the impact of intrinsic fluctuations arising from finite-size effects:
\begin{equation}
\begin{aligned}
\frac{dx(t)}{dt} &= -\alpha x + (1 - x) f(s) + \sqrt{\alpha x + (1 - x) f(s)} \eta_x(t) \\
\frac{dy(t)}{dt} &= -\alpha y + (1 - y) f(s) + \sqrt{\alpha y + (1 - y) f(s)} \eta_y(t)
\end{aligned}
\label{Eq.stoc}
\end{equation}
where $\eta_{x, y}$ are uncorrelated Gaussian white noises, with amplitude $\sigma$ that depends on the network-size $\sigma \propto 1 / \sqrt{N}$ (note that some time dependencies have been omitted for simplicity in the notation).
\begin{figure}[hbtp]
    \centering
    \includegraphics[width=1.0\columnwidth]{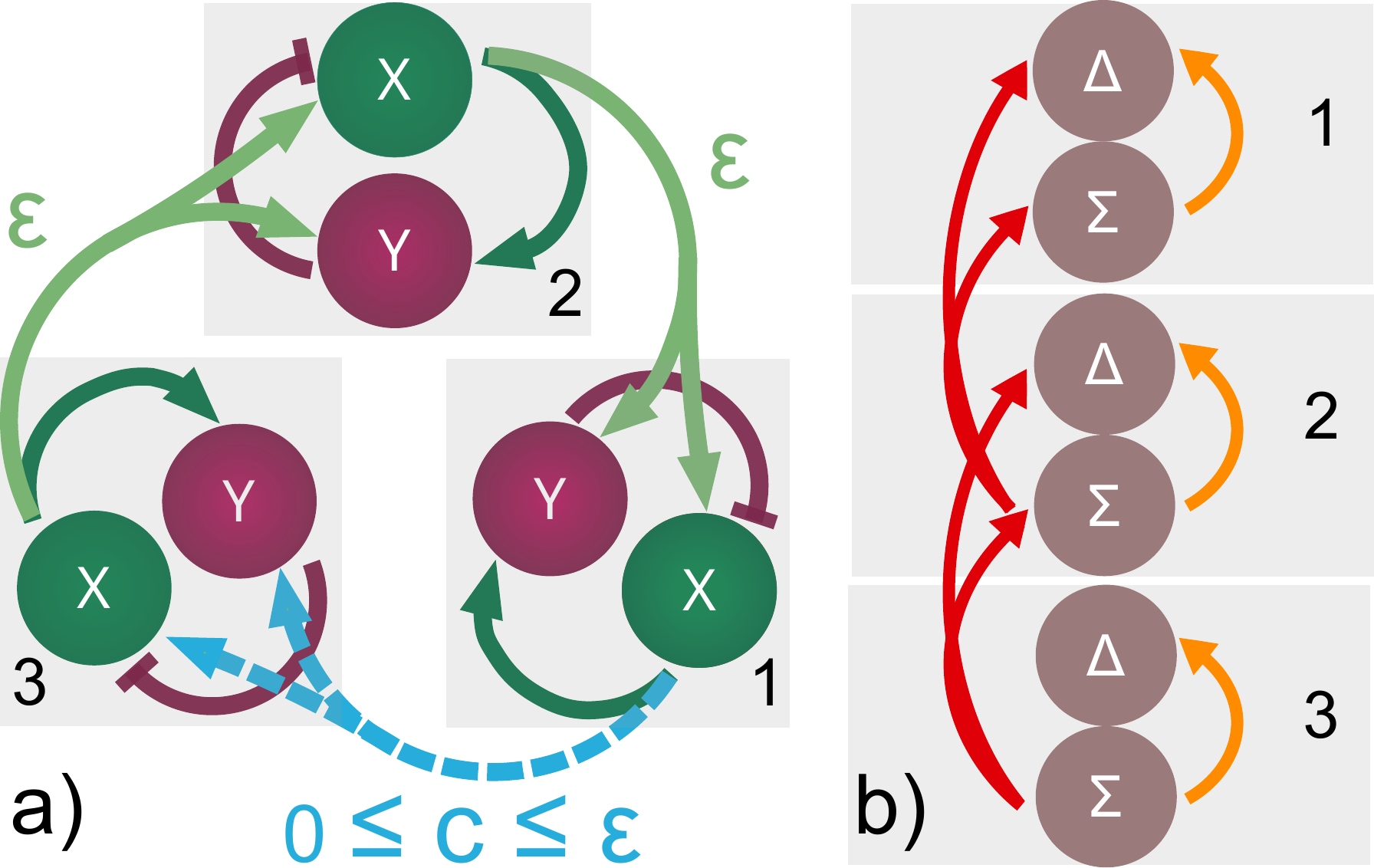}
    \caption{\textbf{Inter-population interactions.} 
    \textbf{(a)}: Schematic representation of the generic non-reciprocal motif. $X$ and $Y$ denote the fractions of active excitatory and inhibitory populations, respectively. Local excitatory couplings are shown in dark green, while inhibitory interactions are indicated in dark red. Light green arrows illustrate inter-population interactions. Two of these connections (represented in light blue) are modulated by the parameter $c$, with $0 \leq c \leq \epsilon$, as defined in Eq.~\eqref{Eq.A}. This modulation allows to interpolate between the homogeneous loop and the feedforward configuration. \textbf{(b)}: Schematic representation of the model in the transformed variables ($\Sigma = x+y$, $\Delta = x-y$). In the transformed variables \textbf{(b)} the feedforward configuration  includes both within-unit (orange) and between-units  (red) couplings, while the cyclic network only includes within-unit (orange) interactions, i.e. the three units are disjoint.}
    \label{Fig.model}
\end{figure}
A key insight  is that even when the deterministic dynamics converge to a non-trivial stable fixed point (i.e., an active state), the presence of non-orthogonal, nearly degenerate eigenvectors can undermine this stability in the presence of noise \cite{Murphy2009, Benayoun2010, DiSanto2018} . Specifically, the poor representation of orthogonal directions in the eigen-basis makes the system highly sensitive to certain perturbations, allowing stochastic fluctuations to push it away from the active state and trap it near the quiescent one. As a consequence, the dynamics becomes intermittent, exhibiting avalanche-like activity: transient, noise-driven excursions of variable amplitude into higher activity states. This illustrates how non-normality, when combined with intrinsic stochasticity, can give rise to rich dynamical regimes that are entirely absent in the purely deterministic system (see \cite{Benayoun2010, DiSanto2018} and Section~\ref{Sec.non-norm} for further discussion).

\vspace{0.25cm}
\subsection{
Structural non-normal effects:
feedforward and feedback network motifs.}

To investigate how non-normal dynamics extend from individual excitatory-inhibitory (x/y) units to interconnected circuits, we analyze simple network motifs composed of multiple coupled x/y populations. Our aim is to understand how the non-normal features identified at the single-unit level are modulated by the architecture of inter-unit interactions.

Here we focus on motifs composed of three x/y populations (or "units" or "areas"), schematically illustrated in Fig.~\ref{Fig.model}, which already capture rich and nontrivial dynamical behavior  (generalization to $n$ coupled populations and more complex motifs are  analyzed 
in the \textit{Supplemental Material}).
Inter-population connectivity is mediated exclusively by excitatory neurons projecting to both excitatory and inhibitory neurons in other populations. In contrast, inhibitory neurons act only locally within each population, in accordance with biological observations~\cite{Kandel2000}. Because inter-population coupling is purely excitatory, the global architecture can be compactly described by a $3\times3$ adjacency matrix $A$, which encodes the excitatory-to-excitatory connectivity pattern:
\begin{equation}
A = \begin{pmatrix}
1 & \epsilon & 0\\
0 & 1 & \epsilon\\
c & 0 & 1
\end{pmatrix}
\label{Eq.A}
\end{equation}
The parameter $0 \leq c \leq \epsilon$ interpolates between a symmetric cycle ($c = \epsilon$) and a purely feedforward structure ($c = 0$), thereby controlling the degree of non-normality. The full $6 \times 6$ adjacency matrix, which includes both excitatory and inhibitory neurons, is not shown explicitly (but note that it is constructed from the previous $3 \times 3$ excitatory submatrix: this submatrix is duplicated in the excitatory-to-inhibitory blocks, while the inhibitory subnetwork remains purely diagonal, reflecting the strictly local nature of inhibition).

The full set of coupled Wilson-Cowan equations that we consider is then:
\begin{equation}
\begin{split}
\dot{\text{x}}_i &= -\alpha \text{x}_i + (1-\text{x}_i)f\bigg(h + A_{i,i}(\gamma_{\mu}\text{x}_i - \gamma_{\nu}\text{y}_i) \\
                 &\hspace{4.8cm} + \gamma_{\ell}\sum_{j \neq i}A_{i,j}\text{x}_j \bigg) \\
\dot{\text{y}}_i &= -\alpha \text{y}_i + (1-\text{y}_i)f\bigg(h + A_{i,i}(\gamma_{\mu}\text{x}_i - \gamma_{\nu}\text{y}_i) \\
&\hspace{4.8cm} + \gamma_{\ell}\sum_{j \neq i}A_{i,j}\text{x}_j \bigg),
\end{split}
\label{Eq.prima}
\end{equation}
with $i=,1, 2$ or $3$ and where $\gamma_\ell$ sets the excitatory coupling strength between units, and $f(s)=\Theta(s)\tanh{s}$, with $\Theta(s)$ the Heaviside function.

 The Jacobian of this system develops complex eigenvalues when three or more coupled units are considered, as detailed in Sec.~\ref{Sec.non-norm}, allowing us to account for interesting transient and time-dependent behaviors, such as noise-induced oscillations~\cite{Fanelli2017}. These $\epsilon$-dependent minimal motifs allow us to isolate how architectural directionality modifies bifurcation structure, eigenvector geometry, reactivity, and noise-driven dynamics, while keeping the local non-normal mechanism fixed. The rest of the paper is organized as follows: in Sec.~\ref{Sec.scaling} we analyze the steady state of Eq.~\eqref{Eq.prima} for the cyclic and feedforward motifs; in Sec.~\ref{Sec.non-norm} we perform a linear stability analysis to characterize the features of the orbits around the stable fixed point; in Sec.~\ref{Subs.Reactivity} we examine the reactivity properties; and in Sec.~\ref{Sec.stochastic} we relate these properties to the behavior of the stochastic system.

\section{Results}
\subsection{Steady state analysis}
\label{Sec.scaling}

First, we analyze the type of bifurcation that the deterministic system undergoes near the vanishing-activity fixed point, i.e., when the system transitions from inhibition-dominated to excitation-dominated dynamics, focusing on the comparison between the two limit architectures introduced above. Subsequently, we study the controllability of the system by quantifying how abruptly the network escapes from the inactive phase, using the excitatory coupling $\gamma_{\mu}$ as a control parameter.

\begin{figure*}[htp]
    \centering
\includegraphics[width=0.8\textwidth]{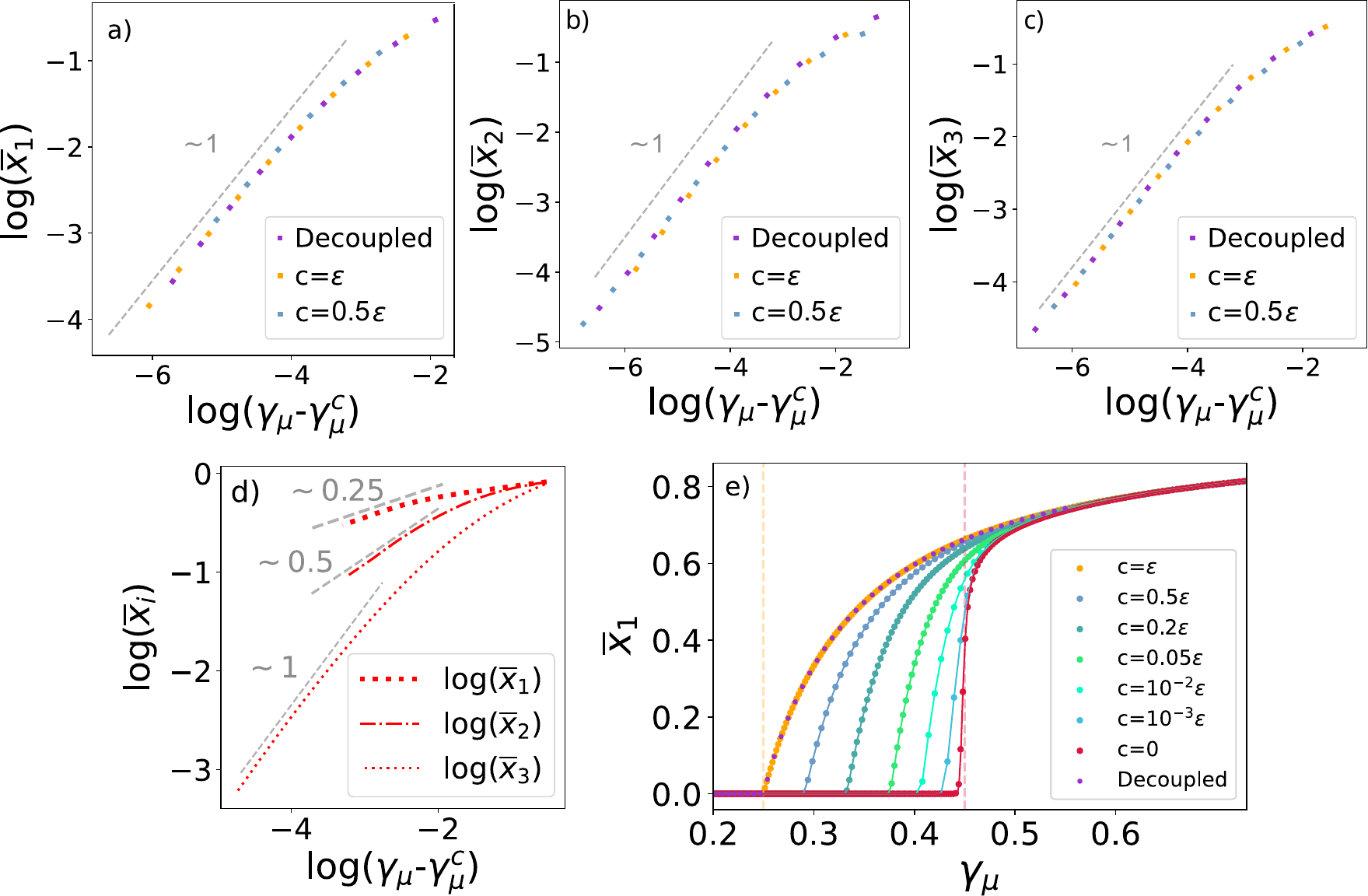}
    \caption{\textbf{Bifurcation diagram and scaling of the fixed points.} 
    \textbf{Panels (a--c)} Scaling of the fixed points around the bifurcation for  $x_1,x_2$, and $x_3$ respectively. The scaling is the same for the decoupled case (purple), for $c = \epsilon/2$ (light blue) and $c = \epsilon$ (orange).
    \textbf{(d)} Scaling of the fixed points around the bifurcation for the Feedforward case ($c = 0$). Note the change of the slope as expected by theoretical results, depending on the population (see legend).
    \textbf{(e)} Bifurcation diagram for the basal population $x_1$, obtained via Runge-Kutta integration. Parameters $\alpha=0.1$, $\gamma_{\nu}=0.35$, and $\gamma_{\ell}=0.2$ are used across several setups. The orange line marks the bifurcation point at $\gamma_{\mu} = \gamma_{\mu}^{c,\epsilon} = 0.25$ for the cyclic system. The decoupled diagram, with bifurcation point predicted at $\gamma_{\mu} = \gamma_{\mu}^{c} = 0.45$, is shifted so that $\gamma_{\mu}^{c,\epsilon} = \gamma_{\mu}^{c}$, resulting in a near-perfect overlap of the cyclic and decoupled curves. Several values of $c$ are also evaluated, up to $c=0$ (feedforward setup), with the expected bifurcation point $\gamma_{\mu}^{c,0} = 0.45$.  }  
    \label{Fig.scaling}
\end{figure*}

For the sake of simplicity, we begin by discussing the decoupled setup ($\gamma_{\ell}=0$), previously studied in \cite{Benayoun2010,DiSanto2018}.  
In this case, the units are trivially copies of the same system, i.e., $x_i = x_j$ and $y_i = y_j$ for all pairs of indices $i,j$ and the equilibrium points satisfy $\bar{x}_i = \bar{y}_i$. We now aim to extend the validity of these equalities to a region surrounding the fixed points. As discussed in detail in Sec.~\ref{Sec.non-norm}, the system is strongly non-normal, and both eigenvectors of the Jacobian align along the bisector of the $x_i$--$y_i$ plane. This alignment allows us to assume that, for $t \gg 1$, $x_i(t) \sim y_i(t)$. Consequently, the dynamical system effectively reduces to a one-dimensional equation. In particular, the normal form around the bifurcation point $\text{x}=0$, with $h = 0$, can be written as
\begin{equation}
    \dot {\text{x}} = (- \alpha + \gamma_{\mu}-\gamma_{\nu})\text{x}- (\gamma_{\mu}-\gamma_{\nu})\text{x}^2 + \mathcal{O}(x^3),
    \label{Eq.deNorm}
\end{equation}
which formally corresponds to the normal form for a transcritical bifurcation \cite{Strogatz2018}. 
The non-zero solution of this equation is
\begin{equation}
   \bar{\text{x}}\sim \frac{ \gamma_{\mu}-\gamma_{\nu}-\alpha}{\gamma_{\mu}-\gamma_{\nu}}.
\end{equation}
and the bifurcation occurs at  the critical value $\gamma_{\mu}^{c}= \gamma_{\nu}+\alpha$. Then, linearizing around this bifurcation point one gets $\bar{\text{x}} \sim \frac{1}{\alpha}\left(\gamma_{\mu}-\gamma_{\mu}^{c}\right)$.

\vspace{0.5cm}
\subsubsection{Cyclic coupling}

In the cyclic architecture ($c = \epsilon$), the system's symmetry ensures that the equations for each node are identical. Consequently, the equilibrium condition $\bar{x}_i = \bar{x}_j$ holds for all pairs of indices $i,j$. Moreover, following a reasoning as the one above, one can show that, as before, excitatory and inhibitory activities also reach the same steady state, $\bar{x}_i = \bar{y}_i$. Therefore, as in the decoupled case, the dynamics can be effectively described by a one-dimensional equation.

The generalization of the decoupled system to the cyclic architecture is straightforward: it is sufficient to replace $\gamma_{\mu}$ with $\gamma_{\mu} + \gamma_{\ell}$. Consequently, the bifurcation point becomes
\begin{equation}
    \gamma_{\mu}^{c,\epsilon} = \gamma_{\nu} + \alpha - \gamma_{\ell},
\end{equation}
and the scaling of the non-zero solution is
\begin{equation}
    \bar{x} \sim \frac{1}{\alpha} \left( \gamma_{\mu} - \gamma_{\mu}^{c,\epsilon} \right),
\end{equation}
i.e., as in the decoupled case, the level of activity grows linearly with the distance to the bifurcation point (thus, the corresponding critical exponent is $\beta=1).$

Fig.~\ref{Fig.scaling}(a) and (b) show the bifurcation diagram for the population $x_1$ and the scaling of the fixed point around the bifurcation point, demonstrating that the curve for the cyclic system as a function of $\gamma_{\mu}$ coincides with that of the decoupled system once appropriately shifted (see figure caption). Moreover, the figure illustrates the scaling of $x_1$ as a function of the distance to the bifurcation point, with an exponent  $\beta=1$.  The figure also illustrates that this scaling exponent holds not only for the cyclic and decoupled systems, but also for a broad range of values $c < \epsilon$, corresponding to architectures that are not perfect cycles, which we refer to as "weak-feedback systems."

These results suggest that cyclic interactions do not fundamentally alter the phase transition of the system, with respect to the uncoupled case, provided that feedback  is preserved ($c>0$).

\subsubsection{Purely feedforward coupling}
In the feedforward system, the units are organized in a chain. We label the populations such that $x_3$ is the upstream or initial node, projecting to $x_2$, which in turn projects to $x_1$. The population $x_3$ behaves as a decoupled unit, becoming active at the bifurcation point $\gamma_\mu = \gamma_\mu^c$, and for $\gamma_\mu \ge \gamma_\mu^c$, activation of $x_3$ triggers a cascade. 

Around the bifurcation point, the dynamics of unit $i$ driven by unit $i+1$ can be expanded as:
\begin{equation}
\begin{aligned}
    \dot {\text{x}_i} &\approx \bar{\mathrm{f}} - \alpha \text{x}_i - \bar{\mathrm{f}}\text{x}_i + \bar{\mathrm{f}}^{\prime}(\gamma_{\mu}-\gamma_{\nu})\text{x}_i \\
    &- \bar{\mathrm{f}}^{\prime}(\gamma_{\mu}-\gamma_{\nu})\text{x}_i^2 -\frac{1}{2}\bar{\mathrm{f}}^{\prime\prime}(\gamma_{\mu}-\gamma_{\nu})^2\text{x}_i^2 + \mathcal{O}(x^3),
\end{aligned}
\label{Eq.feedNorm}
\end{equation}
where $\bar{\mathrm{f}}$ and its derivatives are evaluated for the  averaged external input $s = \epsilon\gamma_{\ell}\bar{\text{x}}_{i+1}$ (except for $i=3$, that has no input). By defining $\delta \equiv \gamma_{\mu}-\gamma_{\mu}^{c}$, we map this to the normal form $\dot{x} = c + ax^2$. The constant driving term scales as $c \propto \bar{\mathrm{f}} \propto \bar{x}_{i+1}$.

Since $x_3$ scales linearly with $\delta$ (standard transcritical scaling), $x_2$ undergoes a saddle-node bifurcation driven by $x_3$, yielding:
\begin{equation}
    \bar{\text{x}}_2\sim \frac{\sqrt{\epsilon\gamma_{\ell}}}{\alpha}\sqrt{\gamma_{\mu}-\gamma_{\mu}^{c}}\sim \delta^{1/2}.
\label{Eq.x2}
\end{equation}
Recursively, $x_1$ is driven by $x_2$, leading to a fourth-root scaling:
\begin{equation}
    \bar{\text{x}}_1\sim \frac{(\epsilon\gamma_{\ell})^{3/4}}{\alpha}\left(\gamma_{\mu}-\gamma_{\mu}^{c}\right)^{1/4} \sim \delta^{1/4}.
\label{Eq.x3}
\end{equation}

Generalizing this recursive argument to a feedforward chain of length $N$, the fixed point of the $k$-th node  in the chain (where $k=1$ is the input and $k=N$ is the output) must scale as:
\begin{equation}
\bar{x}_k \sim (\gamma_{\mu}-\gamma_{\mu}^{c})^{1/2^{k-1}},
\end{equation}
i.e. the exponent $\beta$ depends one the depth of the activated area.
This result implies a rapid "flattening" of the response curve as one moves deeper into the chain. For deep feedforward circuits, the downstream nodes become effectively binary —switching sharply between inactive and active states— thereby acting as robust signal rectifiers despite the linear nature of the initial upstream perturbation.

Fig.~\ref{Fig.scaling}(a/b) show the bifurcation diagram and the scaling of the fixed points in numerical simulations, revealing a perfect agreement with analytical predictions. Comparing the feedforward and cyclic cases, we observe that the fixed point in the feedforward setup is more sensitive to variations in $\gamma_{\mu}$ than in the cyclic architecture.

\subsection{Non-normal dynamics in non-reciprocal motifs}
\label{Sec.non-norm}

Non-normal matrices, i.e. matrices that do not commute with their adjoint (the transpose in the real case), typically have non-orthogonal eigenvectors. In particular, the angle between their right eigenvectors (measured with the standard Euclidean inner product) provides a convenient proxy for non-normality: the smaller the angle, the more pronounced the associated transient effects. Motivated by this, we analyze the eigenvectors of the Jacobian near the equilibrium points and introduce a parameter that quantifies the angle between right eigenvectors. This characterization is key to understanding the phenomenology of the stochastic regime and to highlighting the differences between the cyclic and feedforward architectures.

Let us first rewrite the system in terms of the previously defined variables, $\Sigma$ and $\Delta$
\begin{equation}
\begin{aligned}
    &\Sigma_i = \text{x}_i+\text{y}_i\\
    &\Delta_i = \text{x}_i-\text{y}_i.
\label{Eq.changeCoord}
\end{aligned}
\end{equation}
Since the solution is always of the form $\text{x}_i=\text{y}_i$ $\forall i$, then $\bar{\Delta}_i=0$. Therefore, breaking down $\Sigma_i$ and $\Delta_i$ into the equilibrium and perturbation components, $\Sigma_i = \bar{\Sigma}_i+\sigma_i$ and $\Delta_i = \bar{\Delta}_i+\delta_i$, one can rewrite the linear expansion of the equations as follows:
\begin{equation}
\begin{aligned}
        &\dot \sigma_i = -(\alpha + \bar{\mathrm{f}})\sigma_i + (1-\bar{\Sigma}_i)\bar{\mathrm{f}}^{\prime}\biggl[\sigma_i(\gamma_{\mu}-\gamma_{\nu})+\delta_i(\gamma_{\mu}+\gamma_{\nu})+\\
&\hspace{58mm}+\epsilon\gamma_{\ell}(\sigma_{i-1}+\delta_{i-1})\biggr]\\
        &\dot \delta_i = -(\alpha + \bar{\mathrm{f}}) \delta_i.
\end{aligned}
\label{Eq.primaChanVar}
\end{equation}
Note that, in the new coordinate system, the Jacobian associated with the linearized dynamics assumes an upper-triangular form. This change of basis, originally proposed in \cite{Benayoun2010} for symmetry considerations in the decoupled system, effectively implements the Schur transformation of the Jacobian of Eq.~\eqref{Eq.prima}, guaranteeing its upper-triangular structure in the transformed variables.  

The Jacobian matrices are computed for generic fixed points, as our goal is to derive measures of non-normality applicable across regimes.

\subsubsection{Cyclic couplings}
Let us begin with the study of the cyclic system ($c=\epsilon$). The Jacobian of the cyclic model has a block circulant matrix shape and thus can be block diagonalized according to the circulant matrix theorem \cite{davis1979circulant}, as shown in Supp. Mat. A.1. It is interesting to note that the diagonalization of the blocks leads to the basis $(\Sigma_i,\Delta_i)$. The same result can be achieved by a Schur transformation in the complex field.

A generic circulant matrix is diagonalized by a special unitary matrix known as the discrete Fourier transform matrix, which has $(r, s)$ element equal to $\exp (-2 \pi \mathrm{i}(r-1)(s-1) / n)$. In our model, one can write the diagonalized Jacobian for the cyclic $J_c^{(D)}$
as follows:
\begin{equation}
J_c^{(D)} =
\begin{pmatrix}
\Phi+  \epsilon \Psi  & 0& 0\\
0& \Phi-\epsilon\Psi\biggl(\frac{1}{2}-i\frac{ \sqrt{3}}{2}\biggr) & 0\\
0 &0& \Phi- \epsilon\Psi\biggl(\frac{1}{2}+i\frac{ \sqrt{3}}{2}\biggr)  \\
\end{pmatrix}
\label{Eq.JNOsim}
\end{equation}
where 
\begin{equation}
\begin{aligned}
&\Phi =
\begin{pmatrix}
-\alpha-\bar{\mathrm{f}} + \frac{1}{2}(2-\bar{\Sigma})\bar{\mathrm{f}}^{\prime}(\gamma_{\mu}-\gamma_{\nu}) &  \frac{1}{2}(2-\bar{\Sigma})\bar{\mathrm{f}}^{\prime}(\gamma_{\mu}+\gamma_{\nu})\\
0 & -\alpha -\bar{\mathrm{f}}
\end{pmatrix},
\hspace{3mm}\\
&\Psi =
\begin{pmatrix}
\frac{1}{2}(2-\bar{\Sigma})\bar{\mathrm{f}}^{\prime}\gamma_{\ell} & \frac{1}{2}(2-\bar{\Sigma})\bar{\mathrm{f}}^{\prime}\gamma_{\ell}. \\
0 &0
\end{pmatrix}.
\label{Eq.fi-psi}
\end{aligned}
\end{equation}
Therefore, according to this base, the neural circuit can be remapped into three disjoint feedforward graphs as depicted in Fig. \ref{Fig.model}b.
Using the upper triangular shape of the sub-matrices, we can calculate the eigenvalues in the diagonal elements:
\begin{equation}
    \begin{aligned}
         \lambda_i^{(1)}=&-\alpha-\bar{\mathrm{f}}\\
   \lambda_i^{(2)}= & -\alpha-\bar{\mathrm{f}}+\frac{1}{2}(2-\bar{\Sigma})\bar{\mathrm{f}}^{\prime}(\gamma_{\mu}+k_i\gamma_{\ell}-\gamma_{\nu})
    \end{aligned}
\end{equation}
where $k_i=\biggl\{\epsilon,\epsilon\biggl(-\frac{1}{2}+i\frac{ \sqrt{3}}{2}\biggr),\epsilon\biggl(-\frac{1}{2}-i\frac{ \sqrt{3}}{2}\biggr)\biggr\}$, $i=1,2,3$.
We can identify two complex and four real eigenvalues, wherein global stability is constrained by the inequality $-\alpha-\bar{f} + \frac{1}{2}(2-\bar{\Sigma})\bar{\mathrm{f}}^{\prime}(\gamma_{\mu}+\epsilon\gamma_{\ell}-\gamma_{\nu})<0$. In particular the null fixed point loses its stability when $\gamma_{\mu}= \gamma_{\nu}+\alpha$, as already found in Sec. \ref{Sec.scaling}.
\\

Computing the eigenvectors' elements it becomes evident ----as explicitly shown in in Supp. Mat.  A.1.--- that $\xi = |{\gamma_{\nu}-\gamma_{\mu}-\epsilon\gamma_{\ell}}|/({\gamma_{\mu}+\gamma_{\nu}+\epsilon\gamma_{\ell}}) $ controls the angle between two of the six right eigenvectors. These two eigenvectors encompass contributions of excitatory and inhibitory neurons across all populations; in particular one of them is defined by $\Sigma_i=1,\Delta_i=0$ --i.e. lies in the direction where all excitatory and inhibitory populations have the same firing rate-- and the other one is defined by $\Sigma_i=1,\Delta_i=\xi$. 

Thus, by tuning the parameter $\xi$ one can directly modulate the angle between these two (right) eigenvectors, which determines the degree of non-normality in the system. 

In the asymptotic limit $\xi\to\infty$, the eigenvectors become orthogonal and the system approaches normality. Conversely, for $\xi=0$, $J_c^{(D)}$ becomes defective, as the two eigenvectors coalesce into one. In this regime, non-normality stems from a single structural mechanism governed solely by $\xi$, which therefore uniquely controls the eigenvector collapse. From this perspective, $\xi$ can be viewed as a generalization of the parameter $\zeta = {|\gamma_{\nu}-\gamma_{\mu}|}/({\gamma_{\mu}+\gamma_{\nu}})$, which quantifies non-normality in the decoupled system \cite{DiSanto2018}.
 Furthermore, the degenerate eigenvector defined by $\Sigma_i=1,\Delta_i=0$ extends the corresponding structure found in the decoupled system. Once again, by reabsorbing $\epsilon\gamma_{\ell}$ into $\gamma_{\mu}$ the cyclic system maps onto the decoupled one.

Importantly, this result shows that feedback alone does not generate an independent source of structural non-normality: as long as cyclic symmetry is preserved, the collective dynamics remain reducible to those of a single effective E/I unit. This provides a useful baseline against which genuinely non-reducible architectures, such as feedforward chains, can be identified.

\subsubsection{Feedforward couplings}
We now move on to the feedforward case by setting $c=0$. In this case, the Jacobian matrix takes the form:
\begin{equation}
J_f=
\begin{pmatrix}
\Phi_1 &  \Psi_1\epsilon & 0\\
0& \Phi_2 &  \Psi_2\epsilon\\
0 &0& \Phi_3 \\
\end{pmatrix},
\label{Eq.LinDetFEED}
\end{equation}
where the subscripts indicate that the $2$x$2$ matrices $\Phi_k$ (respectively $\Psi_k$) share the same structure, but are evaluated at different equilibrium points, since each unit of the chain reaches a different steady state. However, as stated above, the equilibrium point of the excitatory and inhibitory sub-populations within each unit, remains homogeneous (i.e. $\bar{\text{x}}_i=\bar{\text{y}}_i$).
The upper diagonal form of $J_f$ allows to read directly the eigenvalues on the diagonal (see Supp. Mat. A.2), i.e.:
\begin{equation}
\{-\alpha-\bar{\mathrm{f}}_i,\frac{1}{2}(2-\bar{\Sigma}_i)\bar{\mathrm{f}}^{\prime}_i(\gamma_{\mu}-\gamma_{\nu})\}, \hspace{2mm} i = 1,2,3.
    \label{Eq.eigenvalFeed}
\end{equation}
Again, the complete form of the eigenvectors, along with their derivation and a generalization to $n$ populations, are provided in Supp. Mat. A.2.

Interestingly, both the elements of the eigenvectors and the eigenvectors themselves exhibit a structured pattern that reflects a recursive-like organization, explicitly highlighting the directionality of signal propagation along the chain. By comparing the pairs of eigenvectors in the Supp. Mat. A.2, two contributions to the system's non-normality can be identified. The first, encoding the so-called \emph{inner} non-normality, is given by $\zeta = |{\gamma_{\nu}-\gamma_{\mu}}|/({\gamma_{\mu}+\gamma_{\nu}})$ which coincides with the non-normality parameter of the decoupled system \cite{DiSanto2018}. Similar to the cyclic setup, $J_f$ becomes defective for $\zeta\rightarrow0$ and normal in the limit $\zeta \rightarrow \infty$. Specifically, if $\zeta = 0$ the eigenvectors in Supp. Mat. A.2., collapse in pairs. The second contribution, representing the \emph{outer} non-normality, is associated with the directionality of the flow along the chain, and coincides with the coupling parameter $$ \gamma_{\ell} $$. 
If, in addition to $\zeta=0$, the inter-layer coupling is strong $(\gamma_\ell \gg \gamma_\mu,\gamma_\nu)$, then all six eigenvectors nearly collapse, and the feedforward architecture yields a dramatic amplification of non-normality.
However, if $\zeta \neq 0$, then no pairs of eigenvectors collapse: $J_f$  cannot become defective by the effect of the directed network architecture alone. 

\begin{figure*}[htp]
    \centering
    \includegraphics[width=0.8\textwidth]{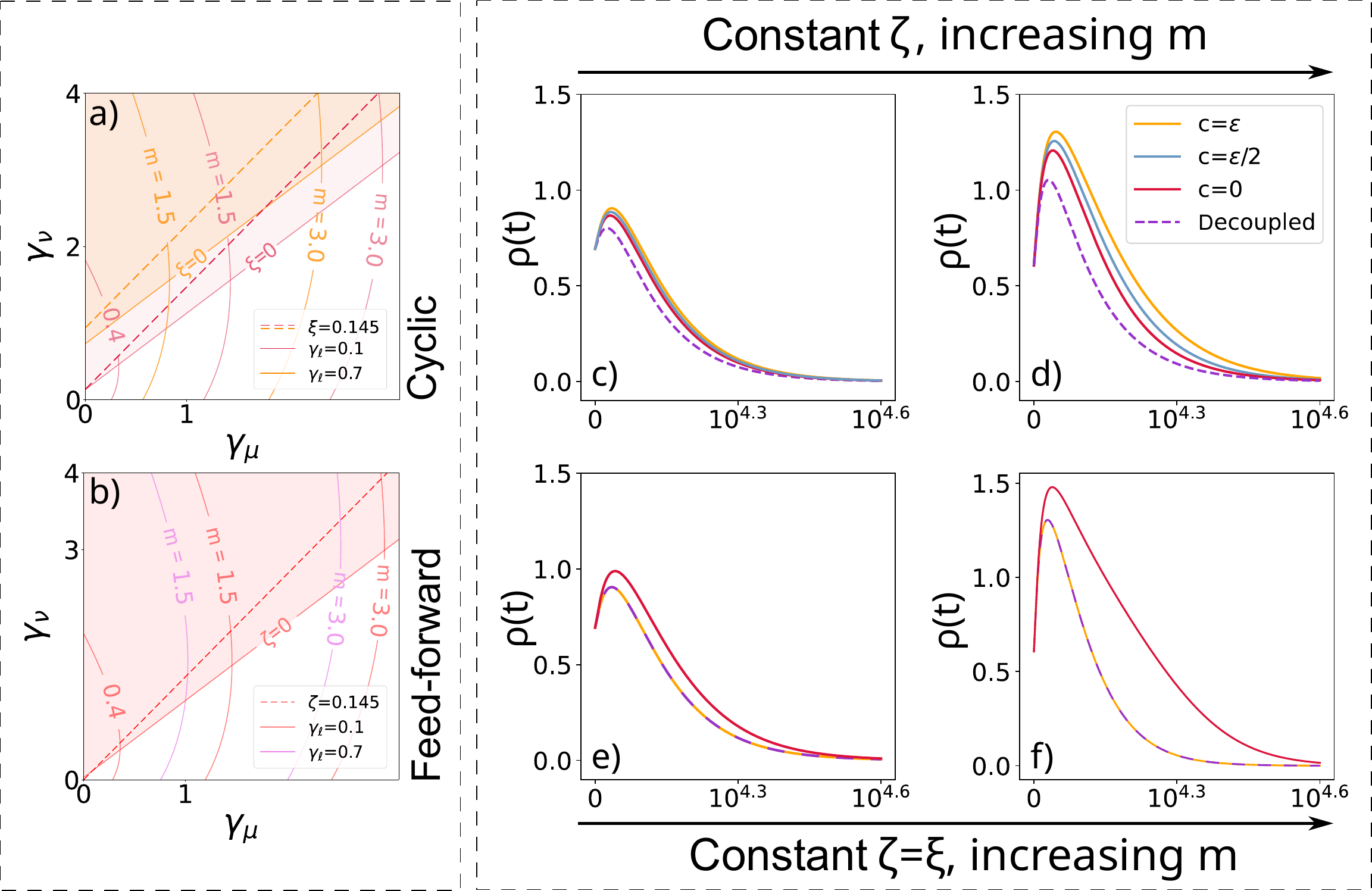}
    \caption{\textbf{Reactivity and comparison of numerical abscissa and non-normality parameters. }\textbf{Left panels. (a,b)}: Level curves of the numerical abscissa $m$ for different values of $\gamma_\ell$, evaluated at the inactive phase. Colored regions indicate the subspace where the level curves are defined. Straight lines correspond to fixed values of the non-normality parameters ($\zeta = \text{const}$ or $\xi = \text{const}$). Panel \textbf{(a)} shows the cyclic setup, while panel \textbf{(b)} shows the feedforward setup. Increasing $\gamma_\ell$ shifts the $m$ level curves leftward, so that $m$ increases for each pair $(\gamma_\nu,\gamma_\mu)$, demonstrating that all possible values of $m$ can be obtained while keeping the non-normality parameter fixed. \textbf{Right panels (c--f)}: Time evolution of $\rho(t)$ under variable reactive conditions. Top row \textbf{(c,d)} corresponds to the decoupled system, with couplings fixed and $\xi$ constant. Bottom row \textbf{(e,f)} corresponds to setups where the non-normality of feedforward and cyclic systems is adjusted to maintain a consistent level across configurations. In this case, only the feedforward system exhibits a significant increase in reactivity. Reactivity increases from left to right with the numerical abscissa $m$.}
    \label{Fig.3}
\end{figure*}

\subsection{Reactivity}\label{Subs.Reactivity}
To quantify the immediate transient amplification capability of the network, we calculate the reactivity $\mathcal{R}$, defined as the maximum eigenvalue of the Hermitian part of the Jacobian \cite{DiSanto2018}:
\begin{equation}
\mathcal{R} = \lambda_{\max}\left(\mathcal{H}(J)\right) = \lambda_{\max}\left(\frac{J + J^{\top}}{2}\right).
\end{equation}
While the spectral abscissa $s(J) = \max_i \text{Re}(\lambda_i)$ determines the asymptotic stability ($t \to \infty$), the reactivity governs the instantaneous growth rate of the perturbation magnitude at $t=0$. This is explicit in the time evolution of the Euclidean norm $\|\delta \mathbf{x}\|^2$:
\begin{equation}
\frac{1}{2}\frac{d}{dt}\|\delta \mathbf{x}\|^2 = \delta \mathbf{x}^{\top} \left(\frac{J + J^{\top}}{2}\right) \delta \mathbf{x} \le \mathcal{R} \|\delta \mathbf{x}\|^2.
\end{equation}
A positive reactivity $\mathcal{R} > 0$ in an asymptotically stable system ($s(J) < 0$) implies a non-monotonic relaxation to the steady state. In the language of critical phenomena, this corresponds to a regime of transient fluctuation amplification: even though the system is formally in the disordered (stable) phase, the violation of detailed balance drives a short-time divergence of the susceptibility, allowing fluctuations to grow macroscopically before the asymptotic restoring forces prevail.

It is crucial to distinguish between non-normality and reactivity. While $\mathcal{R} > 0$ is a sufficient condition for transient growth, non-normality is a necessary but not sufficient condition for reactivity. A system can be non-normal (non-orthogonal eigenvectors) yet non-reactive ($\mathcal{R} < 0$), in which case perturbations decay monotonically but potentially with a relaxation time distinct from the asymptotic inverse gap \cite{Neubert,DiSanto2018}. Thus, $\mathcal{R}$ specifically measures the contribution of the non-symmetric part of the interactions to the local instability.

\subsubsection{Cyclic couplings}
In this case, the numerical abscissa can be evaluated analytically. In particular, we find that for the relevant portion of parameter space  $\gamma_{\mu},\gamma_{\nu}>0$:
\begin{equation}
\begin{aligned}
     m = &\frac{1}{2}(-2\alpha-2\bar{\mathrm{f}} + \frac{1}{2}(2-\bar{\Sigma})\bar{\mathrm{f}}^{\prime}(\gamma_{\mu}+\epsilon\gamma_{\ell}-\gamma_{\nu}))+ \\
     &\frac{1}{2}(2 -\bar{\Sigma})\bar{\mathrm{f}}^{\prime}\sqrt{(\gamma_{\mu}+\epsilon\gamma_{\ell}-\gamma_{\nu})^2+(\gamma_{\mu}+\epsilon\gamma_{\ell}+\gamma_{\nu})^2}\\
\end{aligned}
\label{Eq.m}
\end{equation}
Results are summarized in Fig.\ref{Fig.3}(a) and (b), where we present the level curves of $m$ for several values of $\gamma_{\ell}$ and for $c=0,\epsilon$. Notably, Fig.\ref{Fig.3}(b) shows the behavior of the feedforward system as a function of $m$, which qualitatively matches that of the cyclic case, as shown in Fig.\ref{Fig.3}(a). This observation suggests that Eq.~\eqref{Eq.m} may serve as a qualitative reference even for intermediate configurations with $0<c<\epsilon$.

Interestingly, it turns out that reactivity is governed by two distinct quantities: $\gamma_{\mu}+\epsilon\gamma_{\ell}-\gamma_{\nu}$ and $\gamma_{\mu}+\epsilon\gamma_{\ell}+\gamma_{\nu}$, with the absolute value of their ratio being the non-normality parameter. This implies that, although non-normality is needed to trigger reactivity, a fixed non-normality level does not correspond to a single level of reactivity. In other words, the ratio of $\gamma_{\mu}+\epsilon\gamma_{\ell}-\gamma_{\nu}/\gamma_{\mu}+\epsilon\gamma_{\ell}+\gamma_{\nu}$ can remain constant while adjusting both numerator and denominator to enhance $m$. This is shown explicitly in Fig.\ref{Fig.3}(a) and (b), where the lines representing constant values of $\xi$ or $\zeta$ cross the level curves of $m$, i.e., while moving along a line with constant non-normality, multiple values of reactivity can be achieved.

Moreover, from Eq.~\eqref{Eq.m} we can define  $\Gamma\equiv\gamma_{\mu}+\epsilon\gamma_{\ell}+\gamma_{\nu}$, i.e. the sum of all coupling constants. In what follows we fix the non-normality and distinguish between two regimes: one with weak interactions and weak reactivity ($\Gamma,m\ll1$), which is illustrated in Fig.\ref{Fig.3}(c) and (e) and one with strong interactions and strong reactivity ($\Gamma,m\gg1$), which is illustrated in Fig.\ref{Fig.3} (d) and (f).

Fig.\ref{Fig.3} illustrates the time-evolution of the total activity $\rho(t)$ $=\sqrt{\sum_i (x_i^2+y^2_i)}$, starting from an initial condition close to the bisector, where Eq.~\eqref{Eq.m} holds. In particular, we have chosen parameters allowing the inner systems to be all equivalent, specifically keeping constant the non-normality parameter of the inner systems, $\zeta$. Note the comparison of the evolution of the norm $\rho(t)$ when the reactivity is small (see Fig.\ref{Fig.3}(c)), to a case in which reactivity is large (see Fig.\ref{Fig.3}(d)). As expected, we observe that for large reactivity, the system moves away from the fixed point with vanishing activity before relaxing. From our results, we can conclude that the cyclic system seems to be more reactive than the decoupled system, and that reactivity increases further as $c$ approaches $\epsilon$.

We must note, however, that the reactivity may appear larger for the coupled systems, simply because the coupling itself increases both the reactivity and the non-normality of the system. To isolate the study of the reactivity, we have performed numerical experiments fixing the level of non-normality of the system. In particular, we have chosen parameters in the cyclic system such that
$\xi=\zeta$, i.e., we fix $\gamma_{\mu}$ to compensate the effects of $\gamma_{\ell}$. Therefore, Fig.\ref{Fig.3}\textbf(e) reports the results for low values of reactivity, and Fig.\ref{Fig.3}(f) is for high values of reactivity, showing larger excursions away from the fixed point. We want to emphasize the two main consequences of our analyses: (i) we have verified that the cyclic system can be reduced to the decoupled case merely through coupling parameter shifts (thus, in the cyclic case, non-reciprocity does not create a new effective dynamical degree of freedom beyond the local E/I module) and (ii) we have shown that the feedforward system undergoes a genuine increase in reactivity when internal non-normality is constant. 

\subsubsection{Feedforward couplings}
In contrast, for the feedforward system, there is no unique global parameter to quantify the total non-normality, and simply summing the parameters controlling the inner and outer non-normality would be arbitrary. While a specific choice of couplings could, in principle, make the reactivity coincide, identifying it would require careful fine-tuning, outside of the scope of the present work. Importantly, the presence of an external non-normal component still enhances the system's reactive response. In conclusion, the feedforward network actually increases reactivity compared to the decoupled system. In contrast, in the cyclic system, the enhancement of reactivity can be fully accounted for by matching the level of non-normality. Our results show that feedforward architectures provide a robust way to enhance reactivity without increasing local non-normality, a feature that cannot be achieved in symmetric or cyclic networks.

\subsection{Analysis of the stochastic evolution of the dynamics}\label{Sec.stochastic}

As previously shown in \cite{Benayoun2010,DiSanto2018}, the microscopic underlying system shows very strong finite-size effects. In particular it was shown in \cite{DiSanto2018} using Eq. \eqref{Eq.stoc}, that the demographic noise induces a minimum of the effective potential is the origin, thus generating a bistability in the system. In presence of strong non-normality, the shear flow along the diagonal, associated with the two nearly-collapsing eigenvectors, facilitates the jumps between the two fixed points and the dynamics resembles the avalanching behaviors observed in neural system, where the global activity undergoes large irregular fluctuations between a low-activity and a high-activity state. 

\begin{figure}[htp]
    \centering
    \includegraphics[width=0.5\textwidth]{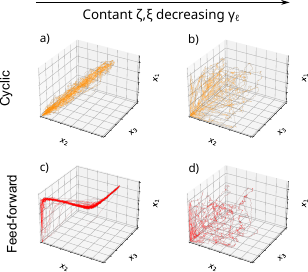}
    \caption{\textbf{Comparison of cyclic and feedforward setups in the stochastic regime.} 
   Top panels show the cyclic configuration; bottom panels show the feedforward setup. Comparing the left and right columns, the system shifts from an incoherent regime (\textbf{b}, \textbf{d}) to either a coherent phase in the cyclic case (\textbf{a}) or step-like trajectories in the feedforward case (\textbf{c}). Parameters: \textbf{(a)} $\gamma_{\ell}=1.13$, $\xi=0.004$; \textbf{(b)} $\gamma_{\ell}=0.13$, $\xi=0.004$; \textbf{(c)} $\gamma_{\ell}=1.13$, $\zeta=0.005$; \textbf{(d)} $\gamma_{\ell}=0.07$, $\zeta=0.005$.
}
    \label{Fig.4}
\end{figure}

Here, we want to explore the behavior of the coupled system under noise effects. We then couple three E/I populations, where each obeys the system size expansion as in \cite{Benayoun2010}:
\begin{eqnarray}
\frac{dx_i}{dt} &= -\alpha x_i + (1 - x_i) f(s_i) + \sqrt{\alpha x_i + (1 - x_i) f(s_i)} \eta_{x_i} \\
\frac{dy_i}{dt} &= -\alpha y_i + (1 - y_i) f(s_i) + \sqrt{\alpha y_i + (1 - y_i) f(s_i)} \eta_{y_i}.
\label{Eq.stocTOT}
\end{eqnarray}
Also in this case, the introduction of noise induces a new stable equilibrium point in zero for all parameter choices, given that $h$ is kept small. Again, when the deterministic stable point bifurcates from zero, bistability appears. Bistability is not a sufficient ingredient to observe avalanching behavior: a high non-normality is needed \cite{DiSanto2018}. However, for too high values of $\xi$ and $\zeta$ the system tends to be trapped in the active minimum. Since we are interested in how the activity propagates across populations, we used the fraction of active excitatory neurons to visualize the trajectories, Fig. \ref{Fig.4}. Similar plots could be obtained for inhibitory activities. 

For low values of the inner non-normality, the system easily gets stuck in its minimum, and excursions between the two minima are not likely.

Once a sufficiently large non-normality level is set, we can distinguish between two separate behaviours, one where the trajectories are incoherent and one where the system shows dynamical patterns, as a function of the parameter $\gamma_{\ell}$. When $\gamma_{\ell}<<\gamma_{\mu},\gamma_{\nu}$ the system lies in an incoherent phase (Fig. \ref{Fig.4} (b) and (d)), while when $\gamma_{\ell}\sim\gamma_{\mu},\gamma_{\nu}$ the behavior is more coherent (Fig. \ref{Fig.4} (a) and (c)). 
When the connectivity among units is large enough and the architecture is cyclic, the activation between units is highly correlated (Fig. \ref{Fig.4}(a)). Instead, in the feedforward case, when the connectivity among units is large enough, the units are activated sequentially. To fix the ideas, let us take two adjacent populations, where the bottom one projects input to --but does not receive any input from-- the top one. When the bottom population transits to the active minimum, it activates also the upper population. The process propagates along the chain, giving rise to the step-like trajectories we observe in Fig. \ref{Fig.4}(c). A discussion about the effective potential induced by the noise and the external couplings is detailed in Supp. Mat. C. 

\section{Discussion and conclusions}

A central motivation of this work, laid out in the Introduction, is that two conceptually distinct sources of non-reciprocity can shape neural dynamics:  local dynamical non-normality  within each E/I module, and  structural non-normality  emerging from directed inter-module connectivity.  Our results clarify when these two ingredients merely “stack” in an effectively reducible way, and when their combination produces genuinely new transient and stochastic behaviors.

At the  local level, each Wilson-Cowan E/I population is tuned to a strongly non-normal regime (captured by the near-alignment of eigenvectors), so that even linearly stable fixed points can display pronounced transient amplification and strong sensitivity to fluctuations.  This “inner” mechanism, quantified by the parameter $\zeta$, sets the baseline susceptibility of each unit to perturbations and noise-driven excursions between inactive and active states.

The  motif or network level (structural) contribution  depends crucially on architecture. In the cyclic (feedback) motif, we find that the effect of coupling can be largely  reabsorbed as an effective shift of the excitatory coupling  (i.e., a renormalization of parameters), so that the deterministic transition and the core local non-normal mechanism are essentially those of the decoupled system in disguise.  In the language of the Introduction, this is a case where directed interactions do not create an independent “structural” non-normal mechanism that fundamentally reshapes the dynamics; instead, the dominant effect is to move the operating point of the same underlying local non-normal module.

By contrast, in the purely feedforward motif, directionality cannot be absorbed into a single effective parameter: the network introduces an additional, explicitly “outer” non-normal component (controlled by $\gamma_\ell$) on top of the inner E/I non-normality $\zeta$.  This distinction becomes especially sharp in the analysis of  reactivity. While non-normality is a prerequisite for reactivity, we show that matching non-normality levels does not uniquely fix reactive amplification; importantly, only the feedforward architecture exhibits a  genuine increase in reactivity when inner non-normality is held constant, consistent with the idea that structural non-normality can add an independent amplification channel.  In other words, this is precisely the kind of “interplay across scales” anticipated in the Introduction: embedding locally non-normal subsystems into a directed architecture can enhance transient responses in a way that is not captured by local properties alone.

These architectural differences are even more pronounced in the stochastic regime, where demographic noise can generate bistability and, with sufficiently strong non-normality, intermittent avalanche-like excursions. Here, structure governs not only the amount of amplification, but also how activity is distributed in time across populations. With strong coupling and feedback (the cyclic motif), noise-driven transitions are typically coherent, producing synchronized activation. In the feedforward motif, strong coupling instead yields sequential, step-like propagation along the chain, reflecting directional recruitment rather than global coherence. Overall, our results distinguish two roles of non-reciprocity: local non-normality permits large noise-driven excursions, while structural asymmetry determines whether those excursions spread synchronously (feedback) or progressively (feedforward).

Our results also connect to classic work on large random interaction matrices in ecology and neuroscience \cite{May1972,SCS,Bunin2017}. In those settings, departing from symmetric (or nearly symmetric) couplings tends to increase eigenvector non-orthogonality and boost transient amplification, making the dynamics much more sensitive to small perturbations. Intuitively, directed (non-reciprocal) interactions can route activity along preferred directions: growth in one subspace can feed into others before relaxation sets in, sustaining irregular fluctuations and, in high dimensions, helping trigger chaotic or strongly time-dependent regimes. From this viewpoint, the “outer” non-normality introduced by directed architecture plays a role analogous to asymmetry in random connectivity: it creates extra pathways for redistributing locally amplified perturbations, hindering simple equilibration and promoting persistent, noise-like dynamics (see also \cite{Martorell2024, Martorell2025}).


Finally, we note limitations and natural extensions. We chose $n=3$ to allow complex eigenvalues and to probe noise-driven time dependence, but we did not observe sustained oscillations; larger $n$ may be needed to obtain more robust quasi-cyclic dynamics. More broadly, an important next step is to move beyond minimal motifs to richer architectures with many more nodes and heterogeneous connectivity, including mixtures of feedforward pathways, recurrent loops, and multiscale feedback, where local and structural non-normal effects could interact in new ways. Testing time-delayed feedback and relaxing the assumption of purely excitatory long-range coupling would further clarify how broadly the separation between inner (local) and outer (structural) non-normality holds in more realistic, large-scale circuits.

Overall, the present framework supports the Introduction’s central claim: non-reciprocity is layered, and understanding neural dynamics requires disentangling when asymmetry acts through local non-normal mechanisms, when it acts through structural non-normal mechanisms, and when their combination yields new transient and noise-driven collective phenomena. We hope this work will encourage further research along these conceptually rich and challenging directions.

\medskip
{\bf{Acknowledgments:}} 
We acknowledge the Spanish Ministry 
and Agencia Estatal de Investigación (AEI), MICIN/AEI/10.13039/501100011033, for financial support, Project PID2023-149174NB-I00 funded also by ERDF/EU.

\def\url#1{}
\bibliography{1-Non-normal}

\begin{thebibliography}{3}%
\makeatletter
\providecommand \@ifxundefined [1]{%
 \@ifx{#1\undefined}
}%
\providecommand \@ifnum [1]{%
 \ifnum #1\expandafter \@firstoftwo
 \else \expandafter \@secondoftwo
 \fi
}%
\providecommand \@ifx [1]{%
 \ifx #1\expandafter \@firstoftwo
 \else \expandafter \@secondoftwo
 \fi
}%
\providecommand \natexlab [1]{#1}%
\providecommand \enquote  [1]{``#1''}%
\providecommand \bibnamefont  [1]{#1}%
\providecommand \bibfnamefont [1]{#1}%
\providecommand \citenamefont [1]{#1}%
\providecommand \href@noop [0]{\@secondoftwo}%
\providecommand \href [0]{\begingroup \@sanitize@url \@href}%
\providecommand \@href[1]{\@@startlink{#1}\@@href}%
\providecommand \@@href[1]{\endgroup#1\@@endlink}%
\providecommand \@sanitize@url [0]{\catcode `\\12\catcode `\$12\catcode `\&12\catcode `\#12\catcode `\^12\catcode `\_12\catcode `\%12\relax}%
\providecommand \@@startlink[1]{}%
\providecommand \@@endlink[0]{}%
\providecommand \url  [0]{\begingroup\@sanitize@url \@url }%
\providecommand \@url [1]{\endgroup\@href {#1}{\urlprefix }}%
\providecommand \urlprefix  [0]{URL }%
\providecommand \Eprint [0]{\href }%
\providecommand \doibase [0]{https://doi.org/}%
\providecommand \selectlanguage [0]{\@gobble}%
\providecommand \bibinfo  [0]{\@secondoftwo}%
\providecommand \bibfield  [0]{\@secondoftwo}%
\providecommand \translation [1]{[#1]}%
\providecommand \BibitemOpen [0]{}%
\providecommand \bibitemStop [0]{}%
\providecommand \bibitemNoStop [0]{.\EOS\space}%
\providecommand \EOS [0]{\spacefactor3000\relax}%
\providecommand \BibitemShut  [1]{\csname bibitem#1\endcsname}%
\let\auto@bib@innerbib\@empty
\bibitem [{\citenamefont {Davis}(1979)}]{davis1979circulant}%
  \BibitemOpen
  \bibfield  {author} {\bibinfo {author} {\bibfnamefont {P.~J.}\ \bibnamefont {Davis}},\ }\href@noop {} {\emph {\bibinfo {title} {Circulant matrices}}}\ (\bibinfo  {publisher} {Wiley, New York},\ \bibinfo {year} {1979})\BibitemShut {NoStop}%
\bibitem [{\citenamefont {Trefethen}(2020)}]{trefethen2020spectra}%
  \BibitemOpen
  \bibfield  {author} {\bibinfo {author} {\bibfnamefont {L.~N.}\ \bibnamefont {Trefethen}},\ }\href@noop {} {\  (\bibinfo {year} {2020})}\BibitemShut {NoStop}%
\bibitem [{\citenamefont {Zakine}\ and\ \citenamefont {Vanden-Eijnden}(2023)}]{zakine2023minimum}%
  \BibitemOpen
  \bibfield  {author} {\bibinfo {author} {\bibfnamefont {R.}~\bibnamefont {Zakine}}\ and\ \bibinfo {author} {\bibfnamefont {E.}~\bibnamefont {Vanden-Eijnden}},\ }\href@noop {} {\bibfield  {journal} {\bibinfo  {journal} {Physical Review X}\ }\textbf {\bibinfo {volume} {13}},\ \bibinfo {pages} {041044} (\bibinfo {year} {2023})}\BibitemShut {NoStop}%
\end{thebibliography}%
\end{document}